\documentstyle[12pt]{article}
\newlength{\abstractwidth}
\def \be {\begin{equation}}
\def \ee {\end{equation}}
\def \bea {\begin{eqnarray}}
\def \eea {\end{eqnarray}}
\def \nn {\nonumber}
\def \ba {\begin{array}}
\def \ea {\end{array}}
\def \la {\langle}
\def \ra {\rangle}
\def \a {\alpha}
\def \b {\beta}
\def \g {\gamma}
\def \G {\Gamma}

\def \lam {\lambda}

\def \s {\sigma}

\def \th {\theta}
\def \Th {\Theta}

\def \dag {\dagger}

\def \cN {{\cal N}}

\begin{document}
\begin{titlepage}

\bigskip
\bigskip\bigskip\bigskip\bigskip
\bigskip \bigskip
\centerline{\Large \bf D-branes as GMS Solitons in Vacuum String Field Theory}

\bigskip\bigskip
\bigskip\bigskip
\centerline{Bin Chen$^{*}$\footnote{chenb@ictp.trieste.it}  and
Feng-Li Lin $^{\dagger}$\footnote{linfl@mail.tku.edu.tw}}
\bigskip
\centerline{\it ${}^*$High Energy Section} \centerline{\it the
Abdus Salam ICTP} \centerline{\it Strada Costiera, 11}
\centerline{\it 34014 Trieste, Italy}
\bigskip
\centerline{\it ${}^\dagger$Department of Physics }
\centerline{\it Tamkang University} \centerline{\it Tamsui, Taipei
25137, Taiwan }
\bigskip\bigskip

\abstract{In this paper we map the D-brane projector states in the
vacuum string field theory to the noncommutative GMS solitons
based on the recently proposed map of Witten's star to Moyal's
star.  We find that the singular geometry conditions of Moore and
Taylor are associated with the commutative modes of these
projector states in our framework. The properties of the candidate
closed string state and the wedge state are also discussed, and
the possibility of the non-GMS soliton in VSFT is commented. }

\end{titlepage}

\setcounter{footnote}{0}

\section{Introduction}
   Vacuum string field theory(VSFT) \cite{RSZ} is a very good tool
to understand the open string tachyon condensation around the
non-perturbative vacuum, around which the BRST operator by
assumption consists only of ghost fields. In this context, the
sliver states and its generalization are algebraically constructed
and identified as the solitonic D25-branes and the lower
dimensional tachyonic Dp-branes \cite{Sen1,Sen2}. Moreover, the
descent relation of the D-brane tension has been verified in the
various cases \cite{Sen1,Okuyama1,Bonora2,CL1}. This is the strong
support to the VSFT ansatz.

  Since the D-branes are the solitonic objects in the VSFT,
it is natural to see if they could be  noncommutative GMS solitons
\cite{GMS} because the Witten's star product of the cubic string
field theory is noncommutative \cite{Witten}. Recently, there is
an explicit identification of Witten's star as Moyal's star
\cite{DLMZ,Bars} so that the cubic string field theory can be
reformulated as an infinite copies of the noncommutative field
theory. (See also \cite{CHL} for a simpler formulation of this map
in the higher energy limit.)

   In the perturbative string theory, the D-brane geometry is
characterized by the boundary condition on the open string endpoints, which
could be Neumann, Dirichlet or the mixing one. On the other hand, in Witten's
string field theory it is the midpoint not the endpoint being selected for the
joining and splitting.  So a pressing question is that: how to characterize the
D-brane geometry in Witten's string field theory? This question is partly
answered in \cite{Sen1} by utilizing the translational invariance. In \cite{MT}
Moore and Taylor provide a sharper distinction between the longitudinal and the
transverse directions to the D-brane by the so-called singular geometry
constraints. These constraints say that the sliver states are the eigenstates
of the half-string momenta and the midpoint coordinates, in a form analogous to
the Neumann and Dirichlet conditions for the endpoints. In this paper we will
see that these eigen-conditions are more transparent in the Moyal language, and
they are characterized by the algebra for the commutative mode.

  In this paper, we will rewrite the sliver state and its higher
rank cousins as the GMS solitons with respect to the Moyal product
and map back to the form with respect to the Witten's star, and
verified that they are indeed the projectors with respect to
Witten's star product.  Our reformulation also gives a more
transparent derivation of the singular geometry conditions of
these projectors than the ones given in \cite{MT} and \cite{Sen2},
the commutative mode play an essential role in yielding the
singular geometry.

  In the next section we summarize the map of Witten's star
to Moyal's star for both the Neumann matrices with and without the
zero mode, especially we emphasize the physical meaning of the
commutative mode. The main results of this paper are contained in
section 3. In section 3.1 we map the sliver state to Witten's star
into the GMS soliton $|0\ra\la 0|$ state to Moyal star. In section
3.2  we discuss the relation between the commutative mode and the
singular geometry and reproduce the eigen-conditions  of Moore and
Taylor. In section 3.3 and 3.4, we generalize the above to the
shifted sliver state and the Dp-brane projector, also their
corresponding singular geometry conditions. In section 3.5, the
higher rank projectors for the Moyal's star are mapped to the one
for the Witten's star, and their properties are discussed. The
details of demonstrating the projector condition is given in the
Appendix. In section 4 we discuss the candidate state for the
closed string based on the ``no-endpoint" algebraic condition. In
section 5, we briefly discuss the wedge state in the Moyal basis.
Finally we conclude our paper in section 6 with few comments,
especially on the possibility of the non-GMS soliton in VSFT.

\section{Map of Witten's star to Moyal's star}
  In \cite{MT} it is found that the pointwise commutative product can be
mapped into Witten's star product for the zero mode of the string
coordinate in the zero slope limit \footnote{In \cite{CL1} we have
generalized this map to the case with constant B-field background
and get the noncommutative product for the zero mode.} with the
following 3-string vertex in terms of the oscillator $a^{\dag}$
\be
\label{v3pw} |V_3\ra =\left(\frac{2}{3\pi}\right)^{1/2}\exp
\sum_{i=1}^3[\frac{1}{6} a^{\dagger }_ia^{\dagger}_i -\frac{2}{3}
a^\dagger_i a^\dagger_{i+1}]|0\ra_{123}. \label{v31}\;,
\ee
where $i\in Z_3$ is the string labels,  so that
\be
\label{map1}
(f\cdot g)(x) =(\la f|\otimes\la g|\otimes\la x|)|V_3\ra \;,
\ee
and $f(x)=\la x|f\ra$, and $\la x|$ is the position state of $x={i\over
\sqrt{2}}(a-a^{\dagger})$, or
\be
\label{xbra} \la x|=\la 0| \exp\{{-1\over2}x^2+i\sqrt{2} a x+
{1\over2} a^2 \}\;.
\ee

   Based on the star spectroscopy \cite{Sen3,Feng}, this map is generalized
to the one for all the higher stringy modes in \cite{DLMZ}, also
in \cite{Belov}, and it turns out that Witten's star product is a
continuous Moyal product with the following correspondence: the
3-string vertex for Witten's star is
\bea
|V_3\ra &=& {\cal N}^{26} \exp\left\{\int_0^{\infty} dk
\sum_{i=1}^3 \left[
-\frac{1}{2}\left(\frac{-4+\th^2}{12+\th^2}\right) (e^{\dag}_i
e^{\dag}_i + o^{\dag}_i o^{\dag}_i)
-\left(\frac{8}{12+\th^2}\right) (e^{\dag}_i e^{\dag}_{i+1} +
o^{\dag}_i o^{\dag}_{i+1})\right.\right.\nn \\
&&\left.\left.  -\left(\frac{4i\th}{12+\th^2}\right) (e^{\dag}_i o^{\dag}_{i+1}
- o^{\dag}_i e^{\dag}_{i+1}) \right]\right\}|0\ra_{123},
\label{stand}
\eea
for pairs of the creation and annihilation operators $(e(k),
e^{\dag}(k), o(k), o^{\dag}(k))$ labelled by the continuous
parameter $0\le k < \infty$, and ${\cal N}$ is some normalization
factor to be fixed later. The Moyal conjugate pairs are
\be
\label{xy} x(k) = \frac{i}{\sqrt{2}}(e(k)-e^{\dag}(k)), \quad
y(k)= \frac{i}{\sqrt{2}}(o(k)-o^{\dag}(k)),
\ee
with the Moyal bracket
\be
\label{Moyal}
[x(k), y(k')]_{\ast} = i\th(k) \delta(k-k')\;,
\ee
where the continuous noncommutativity parameter
\be
\th(k)=2\tanh({\pi k\over 4})\;,
\ee
so that
\be
\label{map2} (f \ast g)(x, y)=(\la x,
y|\otimes \la f|\otimes\la g|)|V_3\ra \;.
\ee
In the above $f(x,y)=\la x, y|f\ra$, and $\la x,y|=\la x| \otimes
\la y|$ with $\la x|$ ($\la y|$) being associated with $x(k)$
($y(k)$). Note that $\th(k=0)=0$ and $\th(k=\pm \infty)=\pm 2$.

 Conventionally
\cite{GJ}, the 3-string vertex is written in the oscillator basis
\be
\hat{X}(\sigma)=\hat{x}_0+\sqrt{2}\sum_{n=1}^{\infty} \hat{x}_n \cos n\sigma
\;, \quad \pi \hat{P}(\sigma)=\hat{p}_0+\sqrt{2}\sum_{n=1}^{\infty} \hat{p}_n
\cos n \sigma\;,
\ee
with the creation and annihilation operators $a_n^{\dag}, a_n$ given by
\be
\label{E}
\hat{x}={i\over 2} E \cdot (a-a^{\dag})\;, \quad \hat{p}=E^{-1} \cdot
(a+a^{\dag})\;, \quad E^{-1}_{mn}=\sqrt{n \over 2}\;\delta_{m,n}+{1\over
\sqrt{b} }\;\delta_{m,0}\delta_{n,0}\;,
\ee
where $b>0$ is an arbitrary parameter.   The above Moyal basis $(e(k), o(k))$
is related to the oscillator basis by
\be
e(k)=\sqrt{2}\sum_{n=0}^{\infty} V_{2n}(k) a_{2n}\;,\quad
o(k)=-i\sqrt{2}\sum_{n=0}^{\infty} V_{2n+1}(k) a_{2n+1}\;,
\ee
with respect to the BPZ conjugate property
\be
bpz\; e(k)= -e^{\dag}(k)\; \quad bpz\; o(k)=-o^{\dag}(k)\;,
\ee
where the vector $V(k)$ labelled by $k$ is the eigenvector of the
Neumann matrices of the 3-string vertex.

  In \cite{DLMZ} the 3-string vertex (\ref{stand}) is the one for the
zero-momentum sector with zero mode excluded, and it is the star product used
to construct the D25-brane sliver \cite{Sen1}. In this case, $V_0(k)\equiv 0$
and
\be
C V(k)=-V(-k)\;, \quad C=(-1)^{m} \delta_{m,n}\;, \quad m,n \ge 1
\ee
where $C$ is the twist operator. From this fact, each eigenvalue is doubly
degenerate for $\pm k$ except at $k=0$ where $e(k=0)=0$ thus $x(k=0)=0$ but
\be
\label{y-mid}
y(k=0)=\frac{i}{\sqrt{2}}(o(k=0)-o^{\dag}(k=0))=-\sqrt{2}\int_0^{\pi/2}
\pi \hat{P}(\sigma) d\sigma =-\sqrt{2} \pi \hat{P}_L\;,
\ee
which is the momentum carried by the left half part of a string \cite{DLMZ}.
Note that $P_L=-P_R$ in the zero-momentum sector.

   Note that the map (\ref{map2}) of the noncommutative products for $k\ne 0$ modes
reduces to the map (\ref{map1}) of the commutative products for
$y(k=0)$ at $k=0$.

   In \cite{Belov} the 3-string vertex of nonzero momentum with zero mode included is also
shown to be in the form of (\ref{stand}) with the same $\th(k)$
but with respect to the different eigenvectors denoted by
$\tilde{V}(k)$ and
\be
\tilde{C}\tilde{V}(k)=\tilde{V}(-k)\;, \quad
\tilde{C}_{m,n}=(-1)^m \delta_{m,n}\;, \quad m,n \ge 0\;,
\ee
so that the nonzero non-degenerate commutative coordinate at $k=0$
is not $y(k=0)$ as in the above but $x(k=0)$. Moreover, it is
shown in \cite{Belov}
\be
\label{x-mid} x(k=0)= \frac{i}{\sqrt{2}}(e(k=0)-e^{\dag}(k=0))=
{1\over \sqrt{2\pi}} \; \hat{X}({\pi\over 2})
\ee
which is the midpoint of the string.

Besides the continuous spectrum labelled by $k$, there exists a
discrete spectrum for the Neumann matrices with zero mode. And it
has been shown in \cite{Belov} the supplemented 3-string vertex
associated with the discrete mode can be also put in the Moyal
form (\ref{stand}), however, with a theta parameter $\th_d(b)\in
(2,\infty)$ determined by the parameter $b$ associated with the
zero mode given in (\ref{E}). In other words, the three-string
vertex is the tensor product of the continuous one with the
discrete one:
\be
|V_3\ra_{p\neq 0}= |V_3\ra_c \otimes |V_3 \ra_d\;.
\ee
Both $|V_3\ra_c$ and $|V_3\ra_d$ could take the Moyal form
(\ref{stand}).


  Before ending this section, we would like to point out that the normalization factor ${\cal N}$
is 1 in the usual SFT
for Witten's star, but for (\ref{map2}) to hold true for the the
Moyal product defined by
\be
(f \ast g)(\vec{x}_3)=\int d\vec{x}_1 d\vec{x}_2 \; K(\vec{x}_1, \vec{x}_2,
\vec{x}_3) f(\vec{x}_1)g(\vec{x}_2)
\ee
with the kernel
\be
K(\vec{x}_1, \vec{x}_2, \vec{x}_3)={1\over \pi^2 \det
\Th}\exp\left(-2i
(\vec{x}_1-\vec{x}_3)\Th^{-1}(\vec{x}_2-\vec{x}_3)\right)\;,
\ee
where $\vec{x}=(x,y)$ and $\Theta=\left( \ba{ccc}
0 &&\th\\
-\th  &&0 \ea \right)$, we need to set\footnote{We will drop the
overall $\pi$ factors from now on, which can be compensated by
proper re-scalings.}
\be
\label{N1} {\cal N}=\exp \{{\log L \over 2\pi} \int^{\infty}_0 dk
\log ({8\over 12+\th^2}) \}\;,
\ee
with the factor ${\log L \over 2\pi}$ as the spectral measure with $L$ the
level regulator, i.e. approximating the infinite $K_1$ matrix by a finite one
of the size $L \times L$\cite{Sen1,HM,DLMZ}. Therefore, the 3-string vertex of
Witten's star is related to the one of Moyal's star by
\be
\label{MW} |V_3\ra_M={\cal N}^{26} |V_3 \ra_W\;,
\ee
and the corresponding projectors are related in an opposite
way.

\section{Sliver state as GMS solitons}
In this section, it  will be shown that the sliver state is a GMS
soliton in the continuous Moyal basis, and the $k=0$ mode gives
the singular geometry in \cite{MT}. Its higher rank cousins will
be constructed and the projector condition with respect to the
3-string vertex of Moyal product will be checked. This establishes
the correspondence between the GMS soliton and the projectors of
VSFT. Some properties of these projectors as multiple D-branes
will also be discussed.

\subsection{Sliver state}
The sliver state of a D25-brane of VSFT constructed in
\cite{KP,Sen1} in the oscillator basis has the form
\be
|\Xi\ra
=[det(1-M)det(1+T)]^{13}\exp(-\frac{1}{2}\sum_{n,m}a^\dagger_m
S_{mn}a^\dagger_n)|p=0\ra
\ee
where
\be
T=\frac{1}{2M}\left(1+M-\sqrt{(1+3M)(1-M)}\right), \hspace{3ex}
S=CT,
\ee
where $C_{mn}=(-1)^m \delta_{m,n}$ is the twist operator, and
$M=CV^{11}$ with $V^{11}$ the one of the Neumann matrices in the
3-string vertex. The eigenvalues of $M$ \cite{Sen3} is
\be
\mu(k)={\th^2-4 \over 12+\th^2}\;,
\ee
with $\th=2\tanh({\pi k\over4})$.


Rewriting the sliver state in the Moyal basis $(e(k), o(k))$ we
have
\be
|\Xi\ra=N^{26} \exp\left( -\frac{1}{2}\int^{\infty}_0 dk \;
\frac{\th-2}{\th+2}(e^\dagger e^\dagger+o^\dagger
o^\dagger)\right) |0\ra\label{sliver2}
\ee
where $N$ is the normalization factor,
\be
\label{N26} \label{N00} N\equiv
\sqrt{det(1-M)det(1+T)}=\exp\left(\frac{\log
L}{2\pi}\int^{\infty}_0 dk \log
\frac{32\th}{(\th+2)(12+\th^2)}\right).
\ee
Note that we have taken into account the double degeneracy of each
eigenvalue at $k\ne 0$ for the normalization factor. The
divergence of the normalization in the limit of $L\rightarrow
\infty$ is quite similar to the discussion in \cite{DLMZ}.

   From the above, we can obtain the string field functional of the
sliver state in terms of the Moyal conjugate variables for $k\ne
0$ modes as
\be
\Psi^{(s)}_W(x(k),y(k))\equiv \la x,y| \Xi\ra = \tilde{N}^{26}
\exp\left(-\int^{\infty}_0 dk\;
\frac{x^2(k)+y^2(k)}{\th(k)}\right)\;, \label{sliver3}
\ee
where
\be
\tilde{N}=\exp\left(\frac{\log L}{2\pi}\int^{\infty}_0 dk \; \log
\frac{16}{12+\th^2} \right).
\ee

The sliver state with respect to Moyal's star is
\be
\Psi^{(s)}_M={\cal N}^{-26} \Psi^{(s)}_W\;,
\ee
which is the continuous version of the GMS soliton of the lowest
rank
\be
2 \;\exp\{-{x^2+y^2 \over \th}\}\;.
\ee

  In summary, the sliver sate obeys the projector condition
for the usual GMS soliton, that is,
\be
(\Psi^{(s)}_{W} \star_W \Psi^{(s)}_{W})(x(k),y(k)) = \Psi^{(s)}_W
(x(k),y(k))\;,
\ee
this also holds if we replace the sub-index $W$ by $M$.  Moreover,
the norm of the sliver state is
\be
\la \Xi|\Xi \ra=\exp\left\{ 26 \frac{\log L}{2\pi}\int^{\infty}_0
dk\; \log {128 \th \over (12+\th^2)^2} \right\}
\ee
which could also be obtained by calculating $\int Dx(k)Dy(k)
|\Psi^{(s)}_W(x(k),y(k))|^2$.

\subsection{Commutative mode and singular geometry}
The algebraic operations in Witten's string field theory are
defined by gluing the left and right half-strings, so the
mid-point is expected to be special because there one loses the
distinction between the left and the right. In \cite{MT} it is
shown that the sliver states obey some singular geometry
conditions with respect to the mid-point.

In the Moyal basis, the mid-point is associated with $k=0$ mode as
implied by (\ref{y-mid}) and (\ref{x-mid}), and this mode is
commutative unlike the other noncommutative $k \ne 0$ modes. It is
easy to see the relation between the singular geometry conditions
and the commutative $k=0$ mode as following. By acting the Moyal
coordinate $y(k)$ defined in (\ref{xy}) on the sliver state
(\ref{sliver2}), we get
\be
y(k')|\Xi\ra =\{N^{26} \exp \int_0^{\infty} dk
\left(-\frac{s_1}{2}(e^\dagger e^\dagger +o^\dagger
o^\dagger)\right)\}\;{i\over \sqrt{2}}[-(s_1(k')+1)
o^{\dagger}(k')]|0\ra\;,
\ee
and the last factor in $[\cdots]$ vanishes only if $k'=0$ since
$s_1(k=0)=-1$. This leads to the result that the sliver state is
the eigenstate of the half-string momentum
$\hat{P}_L=-y(k=0)/(\sqrt{2}\pi)$, that is,
\be
\label{PL0} \hat{P}_L|\Xi\ra =0.
\ee
This is nothing but the singular geometry condition for the
D25-brane as discussed in \cite{MT}. One can translate the above
eigen-condition into the one with respect to the conjugate
coordinate of $\hat{P}_L$, we will not do it here.

  After knowing that the singular geometry is associated with the
$k=0$ commutative mode, we can single out the mode and translate
it into the coordinate basis and examine its behavior in this
basis. However, if we naively continuate the normalization factor
(\ref{N26}) to the $k=0$ mode, we find that it is zero so that we
need to introduce a small cutoff for it, that is, the regularized
$k=0$ projector is
\be
\label{k=0} |\psi_{\epsilon} \ra={\cal
N}^{26}_{\epsilon}\exp\{{1\over2} o^{\dagger}_0o^{\dagger}_0\}|0
\ra
\ee
where $o^{\dagger}_0\equiv o^{\dagger}(k=0)$, and the regularized
normalization constant
\be
{\cal N}_{\epsilon}=\sqrt{4\epsilon \over 3}
\ee
which is the square root of (\ref{N26}) for small $\th=\epsilon$,
the square root is taken because of no double degeneracy as for
the noncommutative modes, i.e. $e(k=0)=0$ as mentioned in section
2.

 From the fact that the integrand of (\ref{sliver3}) is singular at
$k=0$, we know that the transformation of $k=0$ mode in the
$o^{\dag}_0$-basis to the one in the $y(k=0)\equiv y_0$-basis is
singular unless we introduce a small cutoff such that
\be
\label{ry} _{\epsilon_0}\la y_0|=\la 0|
\exp\{{-1\over2}y_0^2+i\sqrt{2} a y_0+ {(1-\epsilon_0)\over2} a^2
\}\;.
\ee
so that
\bea
_{\epsilon_0}\la y_0|\psi_{\epsilon} \ra =({\cal N}_{\epsilon}
{1\over \sqrt{\epsilon_0}})^{26} \exp\{-({1+2\epsilon_0 \over
\epsilon_0}) y_0^2\}  \label{singular1}\;.
\eea
If we choose $\epsilon_0={4\epsilon \over 3}$ and take the limit
of $\epsilon \rightarrow 0$, we obtain
\bea
\lim_{\epsilon \rightarrow 0}\;  {}_{4\epsilon \over 3}\la
y_0|\psi_{\epsilon}\ra=\left\{ \ba{ccc}
0,& & y_0 \neq 0 \\
1,& & y_0= 0 \;.\ea \right. \label{single1}
\eea
This is a projector with unit height due to the choice of the
ratio between $\epsilon$ and $\epsilon_0$. This result is also
discovered by \cite{MT} in the oscillator basis. Note that the
integrand of (\ref{sliver3}) suggests a delta function for $k=0$
mode, this is not case because of the vanishing normalization, and
these two effects match to produce a unit-height function on a
point, which is of zero measure and thus singular in the sense of
functions.

   The most general projector for a commutative coordinate is
the unit-height function with support on discrete points or some
finite intervals. The choice of the commutative projector with
support on finite intervals yields regular mid-point geometry.
However, the above limiting procedure yields only the single-point
support function which implies the zero half-string momentum
condition $P_L=0$. This can be seen as some intrinsic feature of
the sliver state originally derived in the oscillator basis, and
its singular geometry manifests in the commutative mode of the
Moyal basis.

 It might be easy to mix up the singular geometry and the issue of the
midpoint singularity as discussed in  \cite{HM,Sen4}.   Although $k=0$ mode is
associated with both the singular geometry of the mid-point and the midpoint
issue, we shall emphasize that our way of deriving singular geometry condition
is independent of the midpoint singularity issue. The issue concerns how to
consistently extract the finite physical quantities from the infinite
dimensional Neumann coefficient matrix, whose $-1/3$ eigenvalue break the twist
symmetry of the Neumann matrix \cite{Sen3,Feng} and make the physical
observables behave singularly there. The $-1/3$ eigenvalue is exactly the $k=0$
in our formulation. Similar issue happens when one includes the ghost sectors
of the slivers with pure ghost BRST operator inserted at the midpoint
\cite{GW,HM,Sen4,Okuyama1}. On the other hand, the derivation of the
eigen-condition (\ref{PL0}) concerns only the gaussian part of (\ref{sliver2})
and has nothing to do with the twist anomaly or any cut-off of the size of the
Neumann matrix. The irrelevance of the cutoff issue to our derivation can be
also seen from the existence of the well-defined projector (\ref{single1}) in
the limit of $\epsilon \rightarrow 0$. The name of singular geometry is mainly
due to the unit-height function of zero measure for the commutative mode.

\subsection{Shifted D25-brane sliver}
  Instead of starting from the sliver state in the oscillator
basis and then transforming it to the GMS soliton, we can start
with the following general ansatz of the projector in the Moyal
basis
\be
\label{shiftedsliver} |\Xi_s\ra=N_{\parallel}^{26} \exp
\int^{\infty}_0 dk \left(\a e^\dagger + \b o^\dagger
-\frac{s}{2}(e^\dagger e^\dagger +o^\dagger
o^\dagger)\right)|0\ra,
\ee
and require the projector condition
\be
\label{projcon} |\Xi_s\ra_3={_1\la \Xi_s|}{_2\la \Xi_s|}V_3\ra\;.
\ee
Since the inner product is defined with respect to BPZ
conjugation, we have to take $bpz(e^\dagger)=-e,
bpz(o^\dagger)=-o$, so that
\be
\la \Xi_s|=N_{\parallel}^{26}\la 0| \exp \int' dk \left(-\a e - \b
o -\frac{s}{2}(e e + o o)\right)\;.
\ee

After some calculation, the projector condition imposes
the following condition on $s$
\bea
\label{sc} \lefteqn{(\mu_2+(s^{-1}-\mu)^{-1}(\mu_2^2+\mu_3^2))^2-
(\mu_3-2\mu_2\mu_3(s^{-1}-\mu)^{-1})^2} & &\nn\\
&= &\left(s-[\mu+(s^{-1}-\mu)^{-1}(\mu_2^2-\mu_3^2)]\right)
\left(s^{-1}-[\mu+(s^{-1}-\mu)^{-1}(\mu_2^2-\mu_3^2)]\right)
\eea
where
\be
\mu=\frac{\th^2-4}{12+\th^2},\hspace{3ex}
\mu_2=\frac{8}{12+\th^2},\hspace{3ex} \mu_3=\frac{4\th}{12+\th^2}.
\ee

 The solutions of (\ref{sc}) are
\be
s_1=\frac{\th-2}{\th+2}, \hspace{3ex}s_2=\frac{\th+2}{\th-2},
\hspace{3ex} s_3=1. \label{projector}
\ee
This agrees with the 3 solutions found in the oscillator basis
\cite{Sen1}: the $s_1$ solution gives the sliver state as we can
see from the discussion in the last subsection; The $s_2$ solution
gives the non-normalizable projectors as in \cite{Sen1}, which has
the form
\be
T=\frac{1}{2M}(1+M+\sqrt{(1+3M)(1-M)}).
\ee
The functional form of this state is proportional to
$\exp(\frac{x^2+y^2}{2})$, which is non-normalizable so it is
unphysical; the $s_3$ solution gives the identity string field
state.

When $s=s_3=1$, the projection condition on the linear terms
requires that $\a=\b=0$. That means there is no projector state
with a linear term in the exponential acting on the identity
string state. Nevertheless, the states with a linear term in the
exponential acting on the identity string state have been used as
gauge transformations in \cite{Immamura}.

However, when $s=s_1$, there is no constraint on $\a(k)$ and
$\b(k)$ from the projection condition (\ref{projcon}), while the
normalization factor (for Witten's star) is required to be
\be
N_{\parallel}=\exp\{{\log L \over 2\pi} \int^{\infty}_0 dk
\left({\th+2 \over 2}(\a^2+\b^2)+\log {4\th \over
\th+2}\right)\}\;.
\ee
Its string field in the Moyal variables is
\bea
\la x, y|\Xi_s\ra_{s=s_1}&=& \exp\{26{\log L \over 2\pi}
\int^{\infty}_0 dk \; \log 2 \}
\nn \\
& &\exp \int^{\infty}_0 dk \; \frac{-1}{\th} \left\{
\left(x-\frac{i(\th+2)\a}{2\sqrt{2}}\right)^2 +
\left(y-\frac{i(\th+2)\b}{2\sqrt{2}}\right)^2\right\}\;.
\label{GMS}
\eea
This is the shifted GMS soliton with the same norm as the
unshifted one. In the language of the VSFT, this means that the
tension of the shifted sliver is the same as the one for the
unshifted sliver.

   We can obtain the singular geometry condition for the shifted sliver
by acting $y(k)$ on it, the result is
\be
\label{NN} \hat{P}_L|\Xi_s\ra_{s=s_1}= -{i\beta_0\over 2\pi}
|\Xi_s\ra_{s=s_1}\;.
\ee
Alternatively, we can get the singular projector of the
commutative mode as that in (\ref{single1}) but with the support
at $y_0=i\beta_0/\sqrt{2}$. Unlike the singular geometry condition
for the unshifted sliver, it is more clear that (\ref{NN}) is an
eigen-condition because of non-zero eigenvalue.

\subsection{Lower dimensional Dp-brane projector}
  The lower dimensional Dp-brane projectors can be constructed in a
similar way for the sliver as a GMS soliton by using the Moyal
basis for the Neumann matrices with zero mode.  The corresponding
state for the continuous spectrum is the same as the continuous
shifted sliver given by (\ref{shiftedsliver}), we should also
supplement the part for the discrete spectrum with the
noncommutativity parameter $\th_d(b)$.  Explicitly, the Dp-brane
projector is
\bea
|\Xi_s\ra^{(p)}&&=N_{\parallel}^{26} \exp \int' dk \left(\a
e^\dagger + \b o^\dagger -\frac{s}{2}(e^\dagger e^\dagger
+o^\dagger o^\dagger)\right)|0\ra
\nn \\
&&\otimes\; ({32\th_d\over (\th_d+2)(12+\th^2_d)})^{25-p} \exp\left(
-\frac{1}{2}\; \frac{\th_d-2}{\th_d+2}(e_b^\dagger e_b^\dagger+o_b^\dagger
o_b^\dagger)\right) |0\ra\;,
\label{Dp}
\eea
where in the first factor the Moyal pairs $(e^{\dag},o^{\dag})$
along the (25-p)-dimensional transverse directions are related to
the oscillator basis by the eigenvectors of the Neumann matrices
with zero mode, while the rest along the longitudinal directions
to the ones without zero mode.

   The derivation of the singular geometry condition along the
transverse directions is technically the same as the one for the
one of the D25-brane sliver, it gives
\be
\label{DD} \hat{X}^{\perp}({\pi\over2})|\Xi_s\ra^{(p)}_{s=s_1}=
{i\a^{\perp}_0\over 2\sqrt{\pi}}|\Xi_s\ra^{(p)}_{s=s_1}\;.
\ee
Note that $x^{\perp}(k=0)={1\over \sqrt{2\pi}}
\hat{X}^{\perp}({\pi \over 2})$ and $\a^{\perp}_0\equiv
\a^{\perp}(k=0)$, where the superscript $\perp$ denotes the
transverse directions.

Or, we get the following limiting singular projector
\bea
&&\lim_{\epsilon \rightarrow 0}\; {}_{4\epsilon \over 3}\la
x^{\perp}_0|{\cal
N}^{25-p}_{\epsilon}\exp\{\alpha^{\perp}_0e^{\perp
\dagger}_0+{1\over2}\;e^{\perp \dagger}_0e^{\perp \dagger}_0\}|0
\ra =\left\{ \ba{ccc}
0,& & x^{\perp}_0 \neq {i\a^{\perp}_0\over \sqrt{2}} \\
1,& & x^{\perp}_0={i\a^{\perp}_0\over \sqrt{2}} \;,\ea \right.
\label{singularx}
\eea
where the definition of the regularized bra $_{\epsilon_0}\la
x^{\perp}_0|$ is the same as $_{\epsilon_0}\la y_0|$ given in
(\ref{ry}).  These results are also discovered in \cite{MT} for
the D-instanton in the oscillator basis. Here we explicitly
demonstrate that it is associated with the commutative $k=0$ mode.

\subsection{Higher rank projectors and multiple D-branes}
In this subsection we would like to use the map of Witten's star
to Moyal's star to construct the new projector states with respect
to $|V_3\ra$.

We start with the generating function of the projection operators
\cite{Harvey,GMS} on the noncommutative plane defined by its Moyal
pairs. It is
\be
G(\lam, \bar{\lam},x,y)=2e^{-\lam \bar{\lam}+\bar{\lam}
\sqrt{2\over \th}r e^{i\phi}+\lam \sqrt{2\over \th} r
e^{-i\phi}-{r^2 \over \th}}=\sum_{n,m} {\lam^n \over
\sqrt{n!}}{\bar{\lam}^m \over \sqrt{m!}}f_{nm}(x,y)\;,
\ee
where $x+iy=re^{i\phi}$, and $f_{nm}(x,y)$ is the Weyl transform
of the operator $|n \ra \la m|$:
\be
f_{nm}(x,y)=\int dp\; e^{-ipy}\la x+{p\over2}|n \ra \la
m|x-{p\over2}\ra\;.
\ee

 Then we look for the state $|G\ra$ with respect to $|V_3\ra$ for a fixed $k\ne 0$ mode
such that
\be
G(\lam, \bar{\lam}, x,y)=\la x,y| G \ra \;.
\ee
After some manipulations we obtain the projector generating  state
\be
\label{gf} |G \ra = {4 \th \over \th+2} e^{-s \lam
\bar{\lam}-it[\bar{\lam}(e^{\dag}+io^{\dag})+\lam
(e^{\dag}-io^{\dag})]-{1\over2}s(e^{\dagger 2}+o^{\dag
2})}|0\ra\;,
\ee
where
\be
s={\th-2 \over \th+2}\;, \quad t={2\sqrt{\th}\over \th+2}\;.
\ee

Now it is easy to get the projector state for $|n\ra\la n|$ by
differentiating $|G\ra$ $n$ times with respect to $\lam
,\bar{\lam}$ and then set $\lam,\bar{\lam}$ to zero. For example,
the $|0 \ra \la 0|$ corresponds to
\be
\label{00p} |G_{00}\ra={4\th \over \th+2}\;e^{-{1\over2}s(e^{\dag
2}+o^{\dag 2})}|0 \ra \;,
\ee
and the corresponding function form is
\be
f_{00}(x,y)=2e^{-r^2\over \th}\;;
\ee
and the $|1 \ra \la 1|$ corresponds to
\be
|G_{11}\ra = -{4\th \over \th+2}[s+t^2(e^{\dag 2}+o^{\dag
2})]e^{-{1\over2}s(e^{\dag 2}+o^{\dag 2})}|0  \ra \;, \label{g11}
\ee
and its corresponding function form is
\be
f_{11}(x,y)=2(-1+{2\over \th} r^2)e^{-{r^2\over \th}}\;.
\ee
which could be obtained directly from $\la x,y| G_{11}\ra$. One
can explicitly check that (\ref{g11}) is indeed a projector state
with respect to $|V_3\ra$ although it is not in the gaussian form
as the sliver state, the details are given in the Appendix.

   It is straightforward to generalize to the continuous Moyal basis,
for example, the generating projector state is
\be
|G\ra_W= N^{26} \exp\left\{\int^{\infty}_0
dk\left(-s\lam\bar{\lam} -it[\bar{\lam}(e^\dagger+io^\dagger)+
\lam(e^\dagger-io^\dagger)]-\frac{1}{2}s(e^{\dagger 2}
+o^{\dagger 2})\right)\right\}|0\ra \nn \\
\ee
where $N$ is the same normalization factor for the sliver state as
defined in (\ref{N00}), and $s$ and $t$ are now functions of $k$
as understood. Similarly one can then obtain the projectors of the
definite rank in the continuous Moyal basis.

  One can calculate the norm of the $|1\ra\la 1|$ projector,
from $\la G_{11}|G_{11}\ra$ or using the functional integration
over $x(k),y(k)$. It turns out that it is the same as the norm for
the sliver state.

One can also check the orthogonality between the sliver state and
the above projector state by considering their inner product,
which could be defined either on the Moyal basis or in functional
forms. As expected, it is not hard to find that
\be
\la G_{00}| G_{11} \ra=0=\int dx dy f_{00}(x,y)f_{11}(x,y)\;.
\ee

  One may wonder if $|0\ra \la 0|$ and $|1\ra \la 1|$ obey the
same singular geometry condition. This is indeed the case and can
be verified straightforwardly by acting the coordinates $x(k)$ and
$y(k)$ on (\ref{g11}).

   In principle, one could obtain all the projectors from the
projector generating state $|G\ra$, whose ``wavefunctional" is the
generating function of the usual GMS soliton projectors. We expect
the projector states obtained from it have the same norm and
orthogonal to each other. Similarly, the algebraic
eigen-constraints (\ref{NN}) and (\ref{DD})  should also hold for
the higher rank projectors. This implies that we can form the
multiple D-branes state by the linear combinations of these
projectors to form a general GMS soliton. Symbolically that is
\be
O_c=\sum_j c_j |j\ra\la j|\;.
\ee

In the context of VSFT, the string fields are assumed to factorize
into the ghost and matter parts as
\be
\Phi=\Phi_g \otimes \Phi_m \;,
\ee
if it also satisfies
\be
Q \Phi_g+r \Phi_g \ast_g \Phi_g=0\;, \quad Q \Phi_m=0\;,
\ee
where $Q$ is the BRST operator around the closed string vacuum,
then the cubic action of the string field theory can also be
factorize into the ghost and the matter parts, with the matter
action taking the form
\be
\label{maction} S_m = \int \left(\Phi_m \ast \Phi_m - {r\over3}\;
\Phi_m \ast \Phi_m \ast \Phi_m\right)\;.
\ee
Note that we can set $r=1$ by properly re-scaling $\Phi_m$.

If we treat the above Witten's star as a continuous Moyal's star,
then (\ref{maction}) is the action of infinite copies of the
noncommutative field theory for the GMS soliton, where the kinetic
term is suppressed in the large $\theta$ limit \cite{GMS}. On the
other hand, the kinetic term is also dropped here due to the
assumption of VSFT. Follow the discussions in \cite{GMS}, the
absence of the kinetic term makes the action (\ref{maction})
possess a global $U_k(\infty)$ symmetry for each mode labelled by
$k$. Moreover, a level (the number of nonzero $c_j$'s in defining
$O_c$) $m$ soliton spans an infinite dimensional Grassmanian
moduli space parameterized by $\bigotimes_{k\in [0,\infty)}
{U_k(\infty) \over U_k(\infty-m)}$. The partial isometry
\cite{partial} of the Hilbert space for a fixed $k$ relating
different projectors $|j\ra\la j|$ is also the symmetry of the
matter action. This partial isometry could be important when one
tries to take into account the kinetic part of the string field
theory action.

\section{Candidate state for closed string}

  In the last section we see that the stringy midpoint in VSFT
plays the similar role of the zero mode $\hat{x}_0$ in the
perturbative string theory for the Dp-brane's boundary conditions.
It is then natural to expect that the ``stringy endpoints"
$\hat{X}(0)$ and $\hat{X}(\pi)$ in VSFT will play the same role as
in the perturbative string theory. Based on the expectation, a
candidate state $|\G\ra$ for the closed string is proposed in
\cite{MT} by imposing the following ``no-endpoint" algebraic
condition
\be
\label{close0} [\hat{X}(0)-\hat{X}(\pi)]|\G\ra=0 \;.
\ee

   In terms of the Moyal basis, this condition becomes
\be
\label{close1} \int_0^{\infty} dk \; V_o(k) [o(k)-o^{\dag}(k)]
|\G\ra=0\;, \quad V_o(k)\equiv \sum_{n=1}^{\infty} {1\over
\sqrt{2n-1}}V_{2n-1}(k)\;.
\ee
By taking the gaussian ansatz for $|\G\ra$ as
\be
|\G\ra = \exp\left\{-{1\over 2}\int^{\infty}_0 dk dk' \;\G(k,k')
o^{\dag}(k) o^{\dag}(k')\right\}|0\ra \otimes |\G_e\ra \;,
\ee
the ``no-endpoint" condition (\ref{close1}) becomes\footnote{Need
to use $[o(l),\int dk dk'\; {\G(k,k')V_o(k')o^{\dag}(k)+\int dk\;
V_o(k) o^{\dag}(k)} ]=0$ and $\G(k,k')=\G(k',k)$.}
\be
\label{close2} \int_0^{\infty} dk' \;\G(k,k')V_o(k')=-V_o(k)\;,
\ee
while there is no condition on $|\G_e\ra$ which by definition
consists of only $e^{\dag}(k)$'s. It is interesting to note that
we will arrive the same condition (\ref{close2}) even we change
the condition (\ref{close0}) to $\hat{X}(0)-\hat{X}(\pi)=2\pi R$
for the compact direction with radius $R$.

The condition (\ref{close2}) says that $V_o(k)$ is the eigenvector
of $\G(k,k')$ with eigenvalue $-1$. Since $V_o(k)$ is a fixed
vector, this condition alone will not fix a unique $\G(k,k')$.
Moreover, a specific solution $\G(k,k')$  is equivalent to the
ones different by the kernel of $V_o(k)$,
\be
\{K_V\mid \int_0^{\infty} dk' \;K_V(k,k') V_o(k')=0 \;\}\;.
\ee
This suggests that $K_V$ generates the gauge symmetry around the
candidate closed string state, which is similar to the quotient
condition for the BRST constraint. It deserves further study to
see if the gauge symmetry generated by $K_V$ is related to the
gauge symmetry of the closed sting field theory.

In this paper we are not trying to examine the full solution space
of (\ref{close2}) but mention a particular solution
\be
\G(k,k')=-\delta(k,k')\;.
\ee
This gives the above mentioned $s=-1$ state for all modes not just
for $k=0$. As commented in subsection 3.2, $s=-1$ gives the
singular wavefunctional and we get\footnote{Some proper
normalization factor is chosen to arrive the well-defined singular
functional (\ref{close3}).}
\be
\label{close3} \la x(k),y(k)|\G\ra=\left\{ \ba{ccc}
0\;,& & y(k) \neq 0 \quad \forall \;k\\
\la x(k)|\G_e \ra\;,& & y(k) = 0 \quad \forall \;k\ea \right.
\ee
It is not a projector and thus not a solution of the matter sector
in VSFT for arbitrary $\la x(k)|\G_e\ra \ne 1$, and its geometry
is highly singular because it has a single point support for all
modes. This implies that the candidate closed string state is a
highly singular state in VSFT.

\section{Wedge states}
It has been pointed out that the sliver state can be taken as the
limit of wedge state\cite{wedge,Okuyama}. The wedge state is
defined as
\be
|n\ra = (|0\ra)^{n-1}_\star.
\ee
Here the star product is with respect to $|V_3\ra$. As we has seen
in the above sections, the string product could be understood as
the continuous Moyal product. Here, we just assume that the wedge
state is ``diagonal", which means that we define the wedge state
for any fixed k. Let us take the form of the three vertex as in
(\ref{stand}) without integral over $k$, up to a normalization
factor. Then it's not hard to find that the wedge state has the
form
\be
|n\ra
=N_n\exp\left\{-\frac{1}{2}\left(\frac{s+(-s)^{n-1}}{1-(-s)^n}\right)(e^{\dagger
2}+o^{\dagger 2})\right\}|0\ra\;.
\ee

The normalization factor $N_n$ satisfies the recursive relation
\be
N_{n+1}=\frac{1}{1-\mu_1 t_n}N_n \cN
\ee
where $\cN$ is the normalization factor of $|V_3\ra$ and
\be
s=\frac{\th-2}{\th+2}, \hspace{5ex}
t_n=\frac{s+(-s)^{n-1}}{1-(-s)^n}\;.
\ee
Taking the large $n$ limit, if $\th\neq 0$, the exponential term
has exactly the form of the sliver state. This fact support the
original argument \cite{wedge} that the sliver state is taken as
the large $n$ limit of the wedge state. However the normalization
is subtle. From the above relation on normalization factor, it is
easy to notice that the $\cN$ determine the normalization $N_n$.
If we take Witten's vertex, then the normalization factor $N_n$
diverge as $n\rightarrow \infty$. On the contrary, the choice of
Moyal product makes $N_{\infty}$ vanish. It seems that the right
choice is
\be
\cN=\frac{1-s}{1-s+s^2},
\ee
which correspond to the choice in \cite{Okuyama}. However, it is
the form of the projector that we are interested in. One can fix
the normalization of the sliver state with respect to different
$|V_3\ra$'s by projector condition and simple rescaling. But the
form of the exponential part is always the same. In this sense,
the sliver state could be taken as the large $n$ limit of the
wedge states.

One can write down the wavefunction of the wedge state, up to the
normalization factor, it is
\be
\psi^n(x,y)=\la x,y|
n\ra=N_n\frac{1}{1+t_n}\exp\left(-\frac{1}{2}\frac{1-t_n}{1+t_n}(x^2+y^2)\right)\;.
\ee
The generalization to the continuous Moyal product is
straightforward.

\section{Discussions and Conclusions}

   In this paper we map the sliver state and its higher rank cousins
into the GMS solitons in the continuous Moyal basis. Although
their forms in the operator and coordinate bases look different,
they are surprisingly to be the same thing.

We also notice that the midpoint singular geometry of the lump
sliver state is related to projector representation of the $k=0$
commutative algebra, they can be summarized as the following
eigen-conditions with respect to the string midpoint
\be
\label{NNa} \hat{P}^{\parallel}_L|\Xi_s\ra^{(p)}= p_m
|\Xi_s\ra^{(p)}\;.
\ee
and
\be
\label{DDa} \hat{X}^{\perp}({\pi\over2})|\Xi_s\ra^{(p)} =
x_m|\Xi_s\ra^{(p)}\;.
\ee
where $\parallel$ and $\perp$ denote the longitudinal and
transverse directions to the Dp-brane. These conditions are
the analogies to the Neumann and Dirichlet condition for string
endpoints.

In \cite{Bonora2}, it has been argued that the singular property
of the midpoint disappears in the VSFT with NS $B$ field. It is
interesting to understand the smearing feature in the continuous
basis with $B$ field, and see if the projector functional with
constant $B$ field can satisfy some eigen-condition by mixing up
the (\ref{NNa}) and (\ref{DDa}) in analogy to the mixing Neumann
and Dirichlet conditions in the perturbative string theory with
constant $B$ field.

We also find that a specified candidate closed string state defined by the
``no-endpoint" condition is a highly singular state of VSFT with
the gauge symmetry defined by some null condition, and its
geometry is singular for all modes. Its properties deserve further
study to understand the emergence of closed string after tachyon
condensation.

  The matter sector of VSFT in the continuous Moyal basis can
be seen as an continuous tensor product of the non-propagating
noncommutative scalar field theory with noncommutativity $\th(k)$.
So are the sliver and lump projector as the continuous tensor
product of the corresponding GMS solitons for each $k$. For such a
projector it seems that we have the freedom to take different GMS
solitons for different $k$ because the matter action
(\ref{maction}) has an $\bigotimes_{k\in [0,\infty)} U_k(\infty)$
symmetry and a symmetry of the partial isometry. The former
transforms the radially symmetric solution to the general ones
\cite{GMS}, and the latter relates different ranks of the GMS
solitons \cite{partial}. However, just like in the noncommutative
scalar field theory, we expect that the situation could be
different after taking into account the kinetic terms dictated by
the pure ghost BRST operator of the VSFT. It will be important to
analyze the BRST operator and find the classical solution to the
ghost sector in the continuous basis \cite{Belov}. It is also
interesting to investigate Witten's string field from this new
angle.

In the above discussions, we have been focusing on the projectors,
which are just continuous tensor product of the projectors in the
respective Moyal basis. We call such kind of projectors as the
``GMS" projectors since it is nothing but the tensor product of
the usual GMS solitons. It is a very interesting question to find
out the non-GMS projectors, the ones with mixings among different
modes. One simple choice is
\be
|P\ra =N_m \exp\left(-\int dk  dk'\; {\cal M}(k,k')(e^\dag(k)
e^\dag(k')+o^\dag(k) o^\dag(k')) \right)|0 \ra.
\ee
where the mixing function ${\cal M}(k,k')$ will be determined by
the projector condition. However, due to the fact that the Neumann
coefficients are $k$-dependent, the projector condition is too
complicated to be solved.

To by pass the technical difficulty and have some flavor of the
(im)possibility of the non-GMS soliton, we use the fact to be
pointed out in \cite{CHL} that: there is a simple and natural
Moyal basis on the half string $0\le \sigma< \pi/2$ so that the
noncommutativity is $\th=2$ for all stringy modes except for the
commutative string midpoint.

We then try to construct the non-GMS solitons in the $\th=2$ Moyal
basis. The 3-string vertex in this basis is simple
\be
|V_3\ra = \exp\left\{\sum_{i=1,2,3} \int_0^{\pi \over2} d\s \left[
-{1\over2} (e^{\dag}_i(\s) e^{\dag}_{i+1}(\s) + o^{\dag}_i(\s)
o^{\dag}_{i+1}(\s)) -{i\over2} (e^{\dag}_i(\s) o^{\dag}_{i+1}(\s)
- o^{\dag}_i(\s) e^{\dag}_{i+1}(\s)) \right]\right\}|0\ra\;.
\label{stand2}
\ee
Now the noncommutative parameter is the same for any $\s \in
(0,\pi/2]$. In this case, the issue on the non-GMS projector is
workable. We assume the non-GMS projector take the form:
\be
|S\ra=\exp\left\{\int_0^{\frac{\pi}{2}}d\sigma \;
\frac{-s(\sigma)}{2}[(e^\dag(\s)e^\dag(\s)+o^\dag(\s)o^\dag(\s)]
+\int_0^{\frac{\pi}{2}}d\sigma d\s^\prime
x(\sigma,\sigma')(e^\dag(\s)e^\dag(\s^\prime)+o^\dag(\s)o^\dag(\s^\prime))\right\}.
\ee
Since the 3-string vertex takes the same form for all $\s \in
(0,\pi/2]$, the projector condition is far more simple and it
turns out to be
\be
\left\{ \ba{l}
2sx=x \\
\frac{1}{4}(2s+3s^2+3x^2+i(2s-s^2-x^2))=s \ea \right.
\ee
The only solution to the above two equations is $x=0,s=0$, which
is exactly the sliver state solution in $\th=2$ algebra. The
non-GMS projector seems to be impossible in this case. This
suggests that it is hard to find non-GMS projectors in the
continuous Moyal basis of \cite{DLMZ}. More efforts should be made
to have the full answers to the possibility of the non-GMS
soliton.

\vskip 1cm

{\bf Acknowledgements}

FLL thanks Pei-Ming Ho, Jen-Chi Lee and Hyun-Seok Yang for helpful
discussions and also thanks the courtesy of the CTP at NTU, the
NCTS, and the CosPA project, Taiwan. He was supported by the NSC
grant No. NSC89-2112-M-032-002.

\section{Appendix}
In this section, we will outline the step in checking the
projector state (\ref{g11}) with respect to $|V_3\ra$ in the form
(\ref{stand}). The calculation is straightforward although it is
tedious. The relevant formulae have the following form: for
creation and annihilation operator $a^{\dagger}$, $a$ we have
\bea
\lefteqn{\la 0|e^{-\frac{\a}{2}a^2}a^n a^{\dagger
m}e^{-\frac{\b}{2}a^{\dagger 2}
+\gamma a^\dagger}|0\ra}& & \nn\\
&=&(\frac{d}{d\lambda})^n
(\frac{d}{d\mu})^m\left(\frac{1}{\sqrt{1-\a\b}}
\exp\left\{\lambda(1-\a\b)^{-1}(\mu+\gamma)-\frac{1}{2}\lambda^2(1-\a\b)^{-1}\b
\right.\right. \nn \\
& & \left.\left.
-\frac{1}{2}(\mu+\gamma)^2(1-\a\b)^{-1}\a\right\}\right)\mid_{\lambda=\mu=0}
\eea
For instance, we have
\bea
\lefteqn{\la 0|e^{-\frac{\a}{2}a^2} a^{\dagger
}e^{-\frac{\b}{2}a^{\dagger 2}
+\gamma a^\dagger}|0\ra}& & \nn\\
&=&\frac{1}{\sqrt{1-\a\b}}[-\g(1-\a\b)^{-1}\a]\exp(-\frac{1}{2}\g^2(1-\a\b)^{-1}\a) \label{c1}\\
\lefteqn{\la 0|e^{-\frac{\a}{2}a^2}
a^{2}e^{-\frac{\b}{2}a^{\dagger 2}
+\gamma a^\dagger}|0\ra}& & \nn\\
&=&\frac{1}{\sqrt{1-\a\b}}[-(1-\a\b)^{-1}\b+(1-\a\b)^{-2}\g^2]\exp(-\frac{1}{2}\g^2(1-\a\b)^{-1}\a) \\
\lefteqn{\la 0|e^{-\frac{\a}{2}a^2} a^{\dagger
2}e^{-\frac{\b}{2}a^{\dagger 2}
+\gamma a^\dagger}|0\ra}& & \nn\\
&=&\frac{1}{\sqrt{1-\a\b}}[-(1-\a\b)^{-1}\a+\g^2(1-\a\b)^{-2}\a^2]\exp(-\frac{1}{2}\g^2(1-\a\b)^{-1}\a) \\
\lefteqn{\la 0|e^{-\frac{\a}{2}a^2} a^2 a^{\dagger
}e^{-\frac{\b}{2}a^{\dagger 2}
+\gamma a^\dagger}|0\ra}& & \nn\\
&=&\frac{1}{\sqrt{1-\a\b}}[\g(1-\a\b)^{-2}(\a\b+2)-(1-\a\b)^{-3}\g^3\a]\exp(-\frac{1}{2}\g^2(1-\a\b)^{-1}\a) \\
\lefteqn{\la 0|e^{-\frac{\a}{2}a^2}a^2 a^{\dagger
2}e^{-\frac{\b}{2}a^{\dagger 2}
+\gamma a^\dagger}|0\ra}& & \nn\\
&=&\frac{1}{\sqrt{1-\a\b}}[(1-\a\b)^{-2}(\a\b+2)-(1-\a\b)^{-3}\a\g^2(\a\b+5)
+(1-\a\b)^{-4}\a^2\g^4]\nn
\\&&\exp(-\frac{1}{2}\g^2(1-\a\b)^{-1}\a)
\label{c22}
\eea

We would like to prove
\be
|G_{11}\ra \star |G_{11}\ra =|G_{11} \ra
\ee
i.e.
\be
|G_{11} \ra_3={_1\la G_{11}|}{_2\la G_{11}|} |V_3\ra_{123}
\ee
From the contraction between ${_2\la G_{11}|}|V_3\ra$, we obtain
\bea
{_2\la G_{11}|}|V_3\ra&=&{\cal N}
(-\frac{4\th}{\th+2})(1-s\mu_1)^{-1}(c_1+c_2(e^{\dagger
2}_1+o^{\dagger 2}_1)
+d_1e^\dagger_1+d_2o^\dagger_1) \nn\\
& & \cdot \exp\left\{-\frac{1}{2}s(e^{\dagger 2}_1+o^{\dagger
2}_1)-\g_1e^\dagger_1-\g_2o^\dagger_1\right\}
\exp(-\frac{1}{2}s(e^{\dagger 2}_3+o^{\dagger 2}_3))|0\ra_{13}
\eea
where the integration over $k$ is omitted, and
\bea
\g_1&=&\frac{4\th}{(\th+2)^2}(e^\dagger_3-io^\dagger_3) \\
\g_2&=&\frac{4\th}{(\th+2)^2}(o^\dagger_3+ie^\dagger_3) \\
c_1&=&s-2t^2(1-s\mu_1)^{-1}\mu_1+t^2(1-s\mu_1)^{-2}
(\mu_2^2-\mu_3^2)(e^{\dagger 2}_3+o^{\dagger 2}_3)\\
c_2&=&t^2(1-s\mu_1)^{-2}(\mu_2^2-\mu_3^2)\\
d_1&=&t^2(1-s\mu_1)^{-2}[2(\mu_2^2+\mu_3^2)e^\dagger_3+4i\mu_2\mu_3o^\dagger_3]\\
d_2&=&t^2(1-s\mu_1)^{-2}[2(\mu_2^2+\mu_3^2)o^\dagger_3-4i\mu_2\mu_3e^\dagger_3]
\eea
Furthermore, the contraction ${_1\la G_{11}|}{_2\la
G_{11}|}|V_3\ra$ can be obtained from the above relations
(\ref{c1}) to (\ref{c22}). Note that due to the relations
$\g_1^2+\g_2^2=0$, the result could be simplified and the end
result depends only on $c_2$ and the combination
$(\g_1d_1+\g_2d_2)$:
\bea
{_1\la G_{11}|}{_2\la G_{11}|}|V_3\ra&=&{\cal N}
\frac{4\th}{\th+2}(1-s^2)^{-1}(\g_1d_1
+\g_2d_2-2c_2) \nn\\
& &\cdot[s^2(1-s^2)^{-1}-2t^2(1-s^2)^{-2}(s^2+1)]
\exp(-\frac{1}{2}s
(e^{\dagger 2}_3+o^{\dagger 2}_3))|0\ra \nn \\
&=&{\cal N}(\frac{12+\th^2}{8}) |G_{11}\ra_3=|G_{11}\ra_3
\eea
where (\ref{N1}) is used in the last step.


\end{document}